\documentclass[10pt,conference,compsocconf]{IEEEtran}
\usepackage{times}
\usepackage{color}
\usepackage{caption}
\usepackage{cite}
\usepackage{array}
\usepackage{booktabs}
\usepackage{multirow}
\usepackage{url}
\usepackage{flushend}
\usepackage{orcidlink}
\usepackage{url}

\usepackage{breakurl}
\usepackage{hyperref}
\hypersetup{hidelinks,unicode,breaklinks}

\usepackage{ifpdf}
\ifpdf
\pdfoutput=1 
\pdftrue
\fi

\makeatletter
\def\ps@IEEEtitlepagestyle{
	\def\@oddfoot{\mycopyrightnotice}
	\def\@evenfoot{}
}
\def\mycopyrightnotice{
	{\footnotesize
		\begin{minipage}{0.8\textwidth}
			\centering
			Please cite as: Fabian Süß, Marco Freimuth, Andreas Aßmuth, George R. S. Weir, and Bob Duncan, ``Cloud Security and Security Challenges Revisited,'' in \emph{Proc of the 10th International Conference on Cloud Computing, GRIDs, and Virtualization (Cloud~Computing~2019), Venice, Italy}, May 2019.
		\end{minipage}
	}
}
\makeatother
\makeatletter
\let\blx@rerun@biber\relax
\makeatother

\newcolumntype{C}[1]{>{\centering\arraybackslash}p{#1}}

\DeclareRobustCommand*{\IEEEauthorrefmark}[1]{%
	\raisebox{0pt}[0pt][0pt]{\textsuperscript{\footnotesize #1}}%
}

\begin{document}
\pagenumbering{gobble}

\title{\textbf{\Large Cloud Security and Security Challenges Revisited}\\[0.2ex]}

\author{%
	\IEEEauthorblockN{~\\[-0.4ex]\large Fabian S\"u\ss\IEEEauthorrefmark{1}, Marco Freimuth\IEEEauthorrefmark{1}, Andreas A\ss muth\IEEEauthorrefmark{1}\,\orcidlink{0009-0002-2081-2455}, George R S Weir\IEEEauthorrefmark{2} and Bob Duncan\IEEEauthorrefmark{3}\\[0.3ex]\normalsize}
	\IEEEauthorblockA{\IEEEauthorrefmark{1}Technical University of Applied Sciences OTH Amberg-Weiden, Amberg, Germany,\\Email: {\tt \{f.suess $|$ m.freimuth $|$ a.assmuth\}@oth-aw.de}}
	\IEEEauthorblockA{\IEEEauthorrefmark{2}University of Strathclyde, Glasgow, UK, Email: {\tt george.weir@strath.ac.uk}}
	\IEEEauthorblockA{\IEEEauthorrefmark{3}University of Aberdeen, Aberdeen, UK, Email: {\tt robert.duncan@abdn.ac.uk}\\[1ex]}
	\IEEEauthorblockA{\IEEEauthorrefmark{4}}
}

\maketitle

\begin{abstract}
In recent years, Cloud Computing has transformed local businesses and created new business models on the Internet-- and Cloud services are still flourishing. But after the emphatic hype in the early years, a more realistic perception of Cloud services has emerged. One reason for this surely is that today, Cloud Computing is considered as an established and well-accepted technology and no longer as a technical novelty. But the second reason for this assessment might also be numerous security issues that Cloud Computing in general or specific Cloud services have experienced since then. In this paper, we revisit attacks on Cloud services and Cloud-related attack vectors that have been published in recent years. We then consider successful or proposed solutions to cope with these challenges. Based on these findings, we apply a security metric in order to rank all these Cloud-related security challenges concerning their severity. This should assist security professionals to prioritize their efforts toward addressing these issues.
\end{abstract}


\IEEEpeerreviewmaketitle

\section{Introduction}
Cloud security risks have been widely discussed in recent publications by different security specialists, e.g., by the Cloud Security Alliance. \cite{csa2017} Since Cloud services are provided over the Internet, most well-known attacks against web applications are threats against Cloud services, too. In addition, the characteristics of Cloud Computing amplify such existing vulnerabilities and there exist new security challenges that arise from the properties of Cloud services. In order to provide an overview of significant security threats in Cloud Computing, we have grouped the different security issues. In this paper we consider attack vectors against:
\begin{itemize}
	\item Cloud infrastructures,
	\item transportation of data to and from the Cloud and
	\item a client's connection to Cloud services and data.
\end{itemize}
After a brief description of these respective attack vectors in Sections~III to V, we provide some hints and best practice solutions to prevent or reduce the impact of these incidents. In order to help professionals understand the severity of the associated security flaw, we rate these issues using the Common Vulnerability Scoring System (CVSS). \cite{cvss2006}\cite{cvssonline} For those readers who are not familiar with CVSS, a brief introduction is given in Section~II. Having rated the listed attack vectors, we give a final conclusion and outlook in Section~VI.

\section{Common Vulnerability Scoring System}
CVSS is an open framework intended to describe the properties and impact of information security threats and attack vectors. It provides standardised vulnerability scores that are used, for example, in the Common Vulnerability and Exposure (CVE) list or in the National Institute of Standards and Technology's National Vulnerability Database (NVD). To compute a score using CVSS, three metric groups have to be taken into account, Base, Temporal and Environmental, each consisting of a set of metrics. The Base metric group addresses the intrinsic characteristics of a vulnerability that are constant over time and do not depend on specific environments. \textsc{Table}~\ref{tab:base-metrics} shows the two sets of metrics within the Base metric group, the Exploitability metrics and the Impact metrics. In the following sections, the listed abbreviations will be used for the corresponding metrics.

\captionsetup{font={footnotesize,sc},justification=centering,labelsep=period}%
\begin{table}[htbp]
	\caption{Base Metric Group. \cite{cvssonline}}\label{tab:base-metrics}
	\centering%
	\begin{tabular}{ll}
		\toprule
		Exploitability metrics & Impact metrics\\
		\midrule
		Attack Vector (AV) & Confidentiality Impact (C)\\
		Attack Complexity (AC) & Integrity Impact (I)\\
		Privileges Required (PR) & Availability Impact (A)\\
		User Interaction (UI) & \\
		\multicolumn{2}{c}{Scope (S)}\\
		\bottomrule
	\end{tabular}
\end{table}
\captionsetup{font={footnotesize,rm},justification=centering,labelsep=period}%
The Temporal metric group describes time-dependent, but environment-independent factors of a vulnerability and consists of three metrics: Exploit Code Maturity~(E), Remediation Level~(RL) and Report Confidence~(RC) (cf. Table~\ref{tab:temporal-metrics}). 
\captionsetup{font={footnotesize,sc},justification=centering,labelsep=period}%
\begin{table}[htbp]
	\caption{Temporal Metric Group. \cite{cvssonline}}\label{tab:temporal-metrics}
	\centering%
	\begin{tabular}{ll}
		\toprule
		Exploit Code Maturity (E) & Remediation Level (RL)\\
		\multicolumn{2}{c}{Report Confidence (RC)}\\
		\bottomrule
	\end{tabular}
\end{table}
\captionsetup{font={footnotesize,rm},justification=centering,labelsep=period}%
Finally, the Environmental metric group addresses properties of a vulnerability that are relevant and unique to a certain environment. Using these context-specific metrics, scores of the Base metric group or Temporal metric group can be amplified or reduced according to a given setup or scenario. Since we intend to give an overview of Cloud security threats, we do not address environmental specifics and therefore just compute CVSS scores for the Base and Temporal metric groups.\par 
When the metrics are assigned values, the equations of CVSS compute a score ranging from 0.0 (not vulnerable), 0.1 to 3.9 (low), 4.0 to 6.9 (medium), 7.0 to 8.9 (high) and 9.0 to 10.0 (critical). For more detailed information about CVSS, we refer to \cite{cvss2006}\cite{cvssonline}.

\section{Attacks against the Cloud infrastructure}
\subsection{Denial of Service }
A (Distributed) Denial of Service ((D)DoS) attack against a Cloud service provider (CSP) is the cyber criminals attempt to prevent clients from accessing their data or services operated in the Cloud. This can be achieved by flooding the Cloud providers web access with nonsense requests until the servers cannot distinguish between authorised and unauthorised requests and will stop working as a result of the load that leads to the server running out of computing resources. In February 2018, Github was attacked using a so-called memcached DDoS. This means there were no botnets involved, but instead the attackers leveraged the amplification effect of a popular database caching system known as memcached. The attackers flooded memcached servers with spoofed requests and were able to generate incoming traffic for Github at the rate of 1.35 Tbps. \cite{newman_github_ddos} Another variant of a DoS attack could, for instance, be an unplanned shutdown of computing resources due to human error.\par 
DDoS attacks can be started remotely, e.g., over the Internet (AV:N) (cf. \textsc{Table}~\ref{tab:ddos}), without needing special privileges (PR:N) and the complexity of the attack is low (AC:L). No user interaction is needed (UI:N). Usually, DDoS attacks target one specific domain, website or service and therefore we assume the scope of the attack does not cause additional security issues (S:U). In practice, this might not be true for all DDoS attacks and depending on the kind of service under attack, other resources might be affected too. DDoS attacks do not target confidentiality or integrity (C:N, I:N), but availability (A:H).\par
Concerning the temporal metrics, it can be stated that this kind of attack is well-known (RC:C) and attackers know how to perform DDoS attacks (E:H). On the other hand, there are many best practice measures to deal with DDoS attacks (RL:O). 
\captionsetup{font={footnotesize,sc},justification=centering,labelsep=period}%
\begin{table}[htbp]
	\caption{CVSS score for DDoS attacks.}\label{tab:ddos}
	\centering%
	\begin{tabular}{*{5}{C{1cm}}}
		\toprule
		\multicolumn{4}{c}{Base metrics} & Score\\
		\cmidrule{1-4}
		AV:N & AC:L & PR:N & UI:N & \multirow{2}{1cm}{\centering\bf 7.5}\\
		S:U & C:N & I:N & A:H & \\
		\toprule
		\multicolumn{4}{c}{Temporal metrics} & Score\\
		\cmidrule{1-4}
		E:H & RL:O & RC:C & & \bf 7.2\\
		\bottomrule
	\end{tabular}
\end{table}
\captionsetup{font={footnotesize,rm},justification=centering,labelsep=period}%
Specialized firewalls, for example, can detect uncommon behaviour and block the incoming traffic from specific sources. \cite{khadke2016} Newer implementations use techniques of artificial intelligence in order to perform anomaly detection. \cite{alzahrani2018} In addition, load balancing techniques enable CSPs to distribute incoming traffic over different gateways.\par 
This leads to a Base metrics score of 7.5 and a Temporal metrics score of 7.2, which both mean severity ``high''.

\subsection{Malware infection}\label{subsec:malware}
Like any other computer system, Cloud Computing resources are vulnerable to malware infections. The impact of the infection depends on the variety of the malware. For example, an infection by ransomware where data gets encrypted by the attacker would deny users access to their own data, whereas an infection by a keylogger or a rootkit would probably lead to unauthorized access due to stolen credentials.\par
As the Securonix Threat Research Team reported, it only takes minutes before automated attacks against a new exposed IP address begin. The attack itself ranges from attempts to install crypto mining software to irrecoverably deleting databases after the adversary gains access to the system. Both Linux and Windows operating system machines are being attacked. \cite{jaffee_cloud_malware}\par
The most common way to become infected by one of these types of malware is by an unpatched software vulnerability being exploited by an attacker, e.g., when a user opens the malicious attachment of an email (AV:N) that will allow the attacker access (S:C) to the victim's system. The exploit itself can be found easily (AC:L) on special databases like the CVE website. \cite{cve} In order to exploit weaknesses of the targeted system, the malware needs higher privileges (PR:H) and the user also needs to actively click and run the exploit (UI:R). After an attacker has access to a system, integrity (I:H), confidentiality (C:H) and availability (A:H) can no longer be guaranteed.\par
As stated before, an attacker can use pre-built malware kits (E:F) making it easy to generate the malware. Workarounds after patches have been released are usually only temporary because, first of all, they need to be implemented regularly and secondly, they can only fix known issues and are also only available for software under maintenance (RL:W). On the other hand, many of these security issues are demonstrated by security researchers in special scenarios (RC:C).\par
This leads to a high severity, both for the Base (8.4) and the Temporal metrics group (8.1). 
\captionsetup{font={footnotesize,sc},justification=centering,labelsep=period}%
\begin{table}[htbp]
	\caption{CVSS score for malware infection.}\label{tab:malware}
	\centering%
	\begin{tabular}{*{5}{C{1cm}}}
		\toprule
		\multicolumn{4}{c}{Base metrics} & Score\\
		\cmidrule{1-4}
		AV:N & AC:L & PR:H & UI:R & \multirow{2}{1cm}{\centering\bf 8.4}\\
		S:C & C:H & I:H & A:H & \\
		\toprule
		\multicolumn{4}{c}{Temporal metrics} & Score\\
		\cmidrule{1-4}
		E:F & RL:W & RC:C & & \bf 8.1\\
		\bottomrule
	\end{tabular}
\end{table}
\captionsetup{font={footnotesize,rm},justification=centering,labelsep=period}%
There are counter measures against these kinds of attacks like hardening the environment, for example, by rolling out a patch management that keeps the operating system and the application software up to date. Additionally, firewalls and anti-malware software should be the first line of defence to tackle this high threat. Isolation of highly threatened applications, e.g., by using sandboxing, is another efficient countermeasure that should be considered.

\subsection{Unauthorized access}
In this scenario, an attacker or a user has unauthorized access to another user's data or services. This is mostly achieved by stealing (see Subsection~\ref{subsec:malware}, keylogger) or cracking weak passwords. After the adversary has successfully logged in with the user's credentials, he can abuse all the user's services or steal their data. Additionally, more sophisticated attacks like virtual machine escape, when an intruder breaks out of the limitations of a virtual resource in order to get access to other users' resources, are a serious threat due to the necessary wide use of shared resources by virtualisation techniques.\par
To give an example, we refer to the marketing company Exactis which leaked a database containing personal information of users with about 340 million records. \cite{greenberg_database_leak} The incident occurred because the database was publically accessible over the Internet. The relevant web server could easily be found using the search engine `Shodan.' \cite{shodan} Shodan is not specialized in finding web content but systems attached to the Internet according to, for instance, the services these provide. This makes it especially helpful for attackers when searching for vulnerable (Cloud) systems.\par
Since the goal of the attack is to log on remotely with a legitimate user's credentials, the attacker is not limited to physical access to a system (AV:N). The attack complexity on the other hand is considered high (AC:H) because it needs significant technical understanding of the system, e.g., in order to escape a virtual machine's limitations, or psychological skills in order to get credentials using social engineering techniques. As the scope of this attack is to get unauthorized access, it will not be changed (S:U) and user privileges are not required (PR:N) just as user interaction (UI:N). Once an attack provides access to a system, all three security goals, confidentiality (C:H), integrity (I:H) and availability (A:H), can no longer be guaranteed. This is also the reason why this attack has a high base score of 8.1, although it might by very complex to successfully run it.\par
A way to counter this attack on the credential side is to enforce minimum password requirements along with two-factor authentication (RL:W). 
\captionsetup{font={footnotesize,sc},justification=centering,labelsep=period}%
\begin{table}[htbp]
	\caption{CVSS score for unauthorized access.}\label{tab:unauth_access}
	\centering%
	\begin{tabular}{*{5}{C{1cm}}}
		\toprule
		\multicolumn{4}{c}{Base metrics} & Score\\
		\cmidrule{1-4}
		AV:N & AC:H & PR:H & UI:N & \multirow{2}{1cm}{\centering\bf 8.1}\\
		S:U & C:H & I:H & A:H & \\
		\toprule
		\multicolumn{4}{c}{Temporal metrics} & Score\\
		\cmidrule{1-4}
		E:F & RL:W & RC:C & & \bf 7.7\\
		\bottomrule
	\end{tabular}
\end{table}
\captionsetup{font={footnotesize,rm},justification=centering,labelsep=period}%
Additionally, a single user should only have as limited access permissions as needed. Virtual machine escape on the other hand is harder to tackle, and the defence strategy is mainly to have different layers of security enabled, along with an up-to-date patch management.\par   
Encrypting data stored in the Cloud and keeping the decryption key(s) stored on another platform is also a good approach to tackle unauthorized data flow. Exploitations on the social engineering or virtual machine sides have been seen in the past (E:F) and detailed reports are available (RC:C). This lowers the temporal score compared to the base score to 7.7, but this attack scenario is still a serious (high) threat.

\subsection{Data loss}
Data loss describes the event when data is irrecoverably lost by, for example, an environmental catastrophe or a mistaken user interaction. Another scenario in which data might get lost is when data is encrypted but the keys are deleted by accident.\par
Backups should never be accessible over the Internet, so we assume that if an attacker wants to delete data irrecoverably, he needs to have local access to the storage and backup system (AV:L). We assume the attacker does not need special privileges (PR:L). The complexity of the attack is quite low (AC:L), the scope remains unchanged (S:U) and no user interaction is needed (UI:N). Confidentiality is not affected at all (C:N), but integrity (I:H) as well as availability are highly affected as we assume that the data can not be recovered or reconstructed.\par
One of the most catastrophic data losses in recent years was probably the unrecoverable deletion of databases from the popular code managing platform GitLab. \cite{git_data_loss} In early 2017, an engineer wanted to test a new database model for which he set up multiple postgres SQL servers. During the test an abnormally high load occurred causing the new database to stop while performing a backup. For some technical reasons, the backup failed. In the end, only data from a later date could be recovered leading to the loss of recent pulls of about 5,000 projects.\par
In order to deal with data loss, services and data should be operated redundantly, whenever possible not only within the same data centre but also in different locations. A proper backup strategy that includes the testing of all backups is strongly recommended. Since this effectively and reportedly reduces the impact of data loss, we rate this as an ``official fix'' (RL:O). 
\captionsetup{font={footnotesize,sc},justification=centering,labelsep=period}%
\begin{table}[htbp]
	\caption{CVSS score for data loss.}\label{tab:data_loss}
	\centering%
	\begin{tabular}{*{5}{C{1cm}}}
		\toprule
		\multicolumn{4}{c}{Base metrics} & Score\\
		\cmidrule{1-4}
		AV:L & AC:L & PR:L & UI:N & \multirow{2}{1cm}{\centering\bf 7.1}\\
		S:U & C:N & I:H & A:H & \\
		\toprule
		\multicolumn{4}{c}{Temporal metrics} & Score\\
		\cmidrule{1-4}
		E:F & RL:O & RC:C & & \bf 6.6\\
		\bottomrule
	\end{tabular}
\end{table}
\captionsetup{font={footnotesize,rm},justification=centering,labelsep=period}%
There are numerous reports of attacks that lead to data loss (E:F) and most of them can be reproduced easily (RC:C).\par
This assessment leads to a score of 7.1 for the Base metrics group (high) and a temporal score of 6.6 (medium). This emphasises that well-known counter-measures, like storing data redundantly or limiting access to backups, work effectively in practice. The risk of data loss should not be underestimated though.

\section{Attacks on the transportation side}
\subsection{Sniffing / Man in the Middle attacks}
Sniffing means that the adversary is eavesdropping on the communication channel. By observing a client's communication to a Cloud service, the goal of the attacker is to retrieve valuable information. In a so-called Man in the Middle attack, communication from a client to a Cloud service is routed through the attacker. Encrypted traffic might be decrypted by the adversary in order to get the information, re-encrypted and sent to its destination. To give an example, due to misconfiguration leading to allowing public writes to S3 buckets, unauthorised persons could write content to another party's Cloud storage. \cite{olenick_bucket_mitm}\par
Since the attack takes place between the client and the Cloud, the adversary needs access to the adjacent network (AV:A). In cases when communication is not encrypted end to end, this attack scenario is trivial. But even if mechanisms like Transport Layer Security (TLS) are enabled, an attacker is still able to trick (SSL Stripping) the client in order to disable the security efforts. We rate the attack complexity as low (AC:L). Therefore, TLS should always be used along with a technique called HTTP Strict Transport Security (HSTS), which enforces the use of encryption between client and server and makes the attack harder. As stated in the example, a misconfiguration could also lead to a Man in the Middle attack. The attack usually takes place without the user's knowledge (PR:N, UI:N). Depending on the data sniffed, the scope can change and could lead to a break in the confidentiality (C:H), as well as integrity (I:H). 
\captionsetup{font={footnotesize,sc},justification=centering,labelsep=period}%
\begin{table}[htbp]
	\caption{CVSS score for sniffing / Man in the Middle.}\label{tab:mitm}
	\centering%
	\begin{tabular}{*{5}{C{1cm}}}
		\toprule
		\multicolumn{4}{c}{Base metrics} & Score\\
		\cmidrule{1-4}
		AV:A & AC:L & PR:N & UI:N & \multirow{2}{1cm}{\centering\bf 8.1}\\
		S:U & C:H & I:H & A:N & \\
		\toprule
		\multicolumn{4}{c}{Temporal metrics} & Score\\
		\cmidrule{1-4}
		E:F & RL:O & RC:C & & \bf 7.5\\
		\bottomrule
	\end{tabular}
\end{table}
\captionsetup{font={footnotesize,rm},justification=centering,labelsep=period}%
We assume that the attacker wants to stay unnoticed and therefore does not interrupt the communication (A:N). This makes the base score 8.3 and is considered high.\par
There is an offical fix (RL:O) to tackle this attack scenario, namely the use of TLS along with HSTS as described above. But since there is functional exploitation code available (E:F) and detailed reports about this attack exist, the temporal score (7.7) still remains high.

\subsection{Rerouting}
This attack is similar to a Man in the Middle attack. But unlike the attack described above, the attacker usually cannot access plaintext (C:L, I:L) data. The adversary reroutes the packets either on the client or server side (AV:L) in order to prevent successful transmission. The attacker's goal is to make the services operated by the Cloud unavailable (A:H) leading to a Denial of Service. Neither special privileges (PR:N) nor user interaction (UI:N) is needed, but the goal of making the attacked service unavailable remains unchanged (S:U). Since the attacker needs access to the provider's or the customer's gateway, the complexity is considered high (AC:H).\par
As an example, Microsoft's Cloud services were unavailable for several days due to a failure of a public DNS provider, leading to DNS requests targeting Microsoft's Cloud server failing. \cite{ms_dns}
Putting the base values together, a medium value (6.2) shows that the effects of this attack are controllable.\par
Although this attack can easily be discovered (unlike the sniffing approach described above), 
\captionsetup{font={footnotesize,sc},justification=centering,labelsep=period}%
\begin{table}[htbp]
	\caption{CVSS score for rerouting attacks.}\label{tab:rerouting}
	\centering%
	\begin{tabular}{*{5}{C{1cm}}}
		\toprule
		\multicolumn{4}{c}{Base metrics} & Score\\
		\cmidrule{1-4}
		AV:L & AC:H & PR:N & UI:N & \multirow{2}{1cm}{\centering\bf 6.2}\\
		S:U & C:L & I:L & A:H & \\
		\toprule
		\multicolumn{4}{c}{Temporal metrics} & Score\\
		\cmidrule{1-4}
		E:F & RL:W & RC:R & & \bf 5.7\\
		\bottomrule
	\end{tabular}
\end{table}
\captionsetup{font={footnotesize,rm},justification=centering,labelsep=period}%
it can be hard to overcome (RL:W). The best effort would be trying to prevent an attacker from getting access to the local or adjacent network. The attacker has access to a wide range of functional exploitation code (E:F), the attacking vectors are also reasonable (RC:R). This leads to a temporal score of 5.7.

\section{Attacks against the client}\label{sec:client}
\subsection{Malware infection}
As discussed in Subsection~\ref{subsec:malware}, this attack vector is considered on both sides, the client as well as the server. The means attackers use for infection are similiar, and the same applies to potential counter measures. Since the Cloud or the Internet in general offer an easy way for multiple users to work together from all over the world, this benefit can be abused by attackers to infect a lot of clients by successfully attacking a single client (S:C).\par
Reports about upcoming, new malware can be seen almost daily. For example, we consider an Android malware that was reported on 15th June 2018. This malware is specialized as it is multi-functional. It exploits banking details, it stores all characters entered using the on-screen keyboard and has the ability to encrypt the complete device. \cite{malware}\par
The CVSS score is equivalent to the score for malware on the (Cloud) servers. An attacker only needs to send malware via email (AV:N), for example, to infect an end user's device. The attack complexity is quite low (AC:L) as there are pre-built toolkits available on the Darknet. \cite{darknet} 
\captionsetup{font={footnotesize,sc},justification=centering,labelsep=period}%
\begin{table}[htbp]
	\caption{CVSS score for malware infection.}\label{tab:malware-client}
	\centering%
	\begin{tabular}{*{5}{C{1cm}}}
		\toprule
		\multicolumn{4}{c}{Base metrics} & Score\\
		\cmidrule{1-4}
		AV:N & AC:L & PR:H & UI:R & \multirow{2}{1cm}{\centering\bf 8.4}\\
		S:C & C:H & I:H & A:H & \\
		\toprule
		\multicolumn{4}{c}{Temporal metrics} & Score\\
		\cmidrule{1-4}
		E:H & RL:T & RC:C & & \bf 8.1\\
		\bottomrule
	\end{tabular}
\end{table}
\captionsetup{font={footnotesize,rm},justification=centering,labelsep=period}%
A privileged (PR:H) users' action is mandatory (UI:R), the scope can change (S:C) based on the details exposed and all three security goals, confidentiality (C:H), integrity (I:H) and availability (A:H) have to be rated as high, leading to a base score of 8.4.\par
As the example above indicates, functional code exists (E:H) and numerous attacks have been seen in recent years (RC:C). Positively for the user, simple actions are available to deal with this risk at least temporarily (RL:T), like compliance rules for a required minimal security level in order to access Cloud services using a certain client, should be enforced as well as hardening the client's environment. The severity of this threat is still high as a temporal score of 8.1 indicates.

\subsection{Unauthorised data access}
Just as described above, attackers might try to steal user credentials in order to access data without permission. But there are several other examples that might lead to unauthorised access as well. A user might mistakenly be granted higher permissions or a CSP might still process customer data, even after that data has been marked to be deleted by the customer, e.g., in older backups.\par
On 4th September 2018, for instance, an attacker accessed the upload mechanism of the Google Chrome extension ``MEGA''. He added malware code to the extension to steal users' passwords and upload these to the attacker's server. The malware-embedded software was distributed by the normal update process. \cite{mega}\par
The metrics are similar to unauthorised data access on the Cloud side, but limited by the fact that the effects are lower, as a single user usually does not have full access to all the data stored in the Cloud. Additionally, an attacker needs access to the victim's (local) network (AV:L) and data that is stored in the Cloud can be restored more easily (A:L). This leads to a lower base score of 7.7.\par
In addition to that, it is also easier to protect a single client than the complete Cloud infrastructure of a CSP, 
\captionsetup{font={footnotesize,sc},justification=centering,labelsep=period}%
\begin{table}[htbp]
	\caption{CVSS unauthorized data access attacks.}\label{tab:rerouting}
	\centering%
	\begin{tabular}{*{5}{C{1cm}}}
		\toprule
		\multicolumn{4}{c}{Base metrics} & Score\\
		\cmidrule{1-4}
		AV:L & AC:H & PR:L & UI:N & \multirow{2}{1cm}{\centering\bf 7.7}\\
		S:C & C:H & I:H & A:L & \\
		\toprule
		\multicolumn{4}{c}{Temporal metrics} & Score\\
		\cmidrule{1-4}
		E:H & RL:T & RC:C & & \bf 7.4\\
		\bottomrule
	\end{tabular}
\end{table}
\captionsetup{font={footnotesize,rm},justification=centering,labelsep=period}%
although new counter-measures have to be adopted constantly in order to deal with more advanced threats (RL:T). A simple way to reduce the attack surface of such attacks is to enforce a strong password policy or even better two-factor authentication. The maturity of the attack vector, on the other hand, has to be rated higher (E:H) since more malware exists for clients than for Cloud systems. This leads to a temporal score of 7.4.

\section{Conclusion}
Attacks against Cloud infrastructures are multifaceted. They range from Denial of Service attacks to more complex attempts where an attack tries to get unauthorised access. In any discussion about the risk of a Cloud infrastructure, not only the Cloud provider's side should be considered but also the transportation of data as well as the security of the endpoints connected to the services and data operated by the Cloud. The CVSS scoring helps companies to identify the most critical security flaws. Based on the attack vectors and vulnerabilities described in the previous sections, we used the Temporal metrics score to rank these security challenges (cf. Table~\ref{tab:ranking}). 
\captionsetup{font={footnotesize,sc},justification=centering,labelsep=period}%
\begin{table}[htbp]
	\caption{Ranking of the most sever Cloud security challenges.}\label{tab:ranking}
	\centering%
	\begin{tabular}{clc}
		\toprule
		\bf Rank & \bf Security challenge & \bf Score\\
		\midrule
		1 & Malware infection (Cloud infrastructure) & 8.1\\
		2 & Unauthorised access & 7.7\\
		3 & Man in the Middle attacks & 7.5\\
		4 & DDoS attacks & 7.2\\
		5 & Data loss & 6.6\\
		6 & Rerouting & 5.7\\
		\bottomrule
	\end{tabular}
\end{table}
\captionsetup{font={footnotesize,rm},justification=centering,labelsep=period}%
This table only contains security problems for the Cloud infrastructure and the data connection between clients and the Cloud services. These are the aspects a CSP has under their sole control and from a customer's perspective, these are the most important properties that should be addressed when a contract with a CSP is negotiated. In a more specialised setting, e.g., for a Cloud-based SCADA (supervisory control and data acquisition) system, the situation might be easier because of the smaller number of potentially different clients.\par
Vulnerabilities or attacks targeting specifically the client side are hard to deal with for most CSPs. The clients are usually not managed or controlled by the CSP but by the customers themselves. If a CSP offers a multi purpose Cloud service that should be usable by any customer and any device, it is hard to deal with all potential vulnerabilities of all possible combinations of client applications and operating systems. Nevertheless, it is highly recommended not to support ``ancient'' client software (applications as well as operating systems) for which security updates have officially been discontinued. However, the security challenges for clients described in Section~\ref{sec:client}, malware infections (8.1) and unauthorised data access (7.4), both have to be considered as highly severe. Therefore, customers are advised to install security patches as soon as those are available and also to have a proper strategy for access control.\par 
In order to obtain a scoring of the discussed security challenges for a unique environment, we suggest to add the metrics of the Environmental metric group according to the given scenario. On the Internet, there are several CVSS~3.0 calculators available, like the one provided by the Forum of Incident Response and Security Teams (FIRST) \cite{cvss3calc}), that can be used to do the calculations easily.

\section*{Acknowledgment}
The authors would like to thank the Bavarian Research Alliance (BayFOR) for funding several visits of the partners involved in this paper. This funding definitely helped to develop joint research and teaching activities.

\end{document}